\documentclass[twocolumn, superscriptaddress,
 amsmath,amssymb,
 aps,prl,
]{revtex4-2}

\usepackage{graphicx}
\usepackage{dcolumn}
\usepackage{bm}
\usepackage{physics}
\usepackage{here}
\usepackage{xcolor}
\usepackage[
 colorlinks=true,
 allcolors=blue,
]{hyperref}
\newcommand{\red}[1]{{\textcolor{black}{#1}}}

\begin{document}

\preprint{APS/123-QED}

\title{Gapless superconductivity and its real-space topology in quasicrystals}
\author{Kazuma Saito}
\affiliation{Department of Applied Physics, Tokyo University of Science, Katsushika, Tokyo 125-8585, Japan}
 
\author{Masahiro Hori}
\affiliation{Department of Applied Physics, Tokyo University of Science, Katsushika, Tokyo 125-8585, Japan}%
\affiliation{%
Department of Physics and Engineering Physics, and Centre for Quantum Topology and Its Applications (quanTA), University of Saskatchewan, 116 Science Place, Saskatoon, Saskatchewan, Canada S7N 5E2
}%

\author{Ryo Okugawa}
\affiliation{Department of Applied Physics, Tokyo University of Science, Katsushika, Tokyo 125-8585, Japan}%

\author{K. Tanaka}
\affiliation{%
Department of Physics and Engineering Physics, and Centre for Quantum Topology and Its Applications (quanTA), University of Saskatchewan, 116 Science Place, Saskatoon, Saskatchewan, Canada S7N 5E2
}%

\author{Takami Tohyama}
\affiliation{Department of Applied Physics, Tokyo University of Science, Katsushika, Tokyo 125-8585, Japan}%

\date{\today}

\begin{abstract}
We study superconductivity in Ammann-Beenker quasicrystals under magnetic field. By assuming an intrinsic $s$-wave pairing interaction and solving for mean-field equations self-consistently, 
we find gapless superconductivity in the quasicrystals at and near half filling.
We show that gapless superconductivity \red{results from the combination of} 
broken translational symmetry and confined states \red{that is characteristic of} 
the quasicrystals.
When Rashba spin-orbit coupling is present, 
the quasicrystalline gapless superconductor can be topologically nontrivial and 
characterized by a nonzero pseudospectrum invariant given by a spectral localizer.
The gapless topological superconducting phase exhibits edge states with near-zero energy. 
These findings suggest that quasicrystals can be a unique platform for realizing gapless superconductivity with nontrivial topology.
\end{abstract}

\maketitle
\textit{Introduction.}
Quasicrystals (QCs) are materials with aperiodic structure that has a long-range order \cite{Shechtman_1984, Levine86}.
Since translational symmetry is absent, one cannot employ a momentum-space picture to understand quantum phenomena in QCs.
One such example is superconductivity, as the concept of Fermi surfaces in momentum space cannot be applied directly to QCs
\cite{Araujo19, Sakai17, Sakai19, Takemori20}.
Nevertheless, superconductivity has recently been discovered in some QCs experimentally \cite{Kamiya_2018,Tokumoto24, Terashima24arXiv}.

There has recently been theoretical works showing that
not only conventional superconductivity, but also topologically nontrivial superconductivity can occur in QCs \cite{Fan21, Fulga16, Loring19, Varjas19, Cao20, Hua21, Ghadimi21, Wang22, Liu23, Manna24, Qi24, Hori24arXiv, Hori24arXiv2}.
While unconventional pairing symmetry such as $p$-wave is typically required to realize topological superconductivity, even with $s$-wave pairing,
spin-orbit-coupled superconductors can be topologically nontrivial under strong magnetic field \cite{Sato09, Sato10}. 
It has been shown theoretically by some of the present authors that such
topological superconductivity with broken time-reversal symmetry is realizable even in QCs when the chemical potential is near the top or bottom of the kinetic-energy band \cite{Ghadimi21, Hori24arXiv}.
At the same time, however, there is no fundamental difference in the nature of topological superconductivity compared with that in periodic systems and thus, no new phenomenon that is \red{specific} %
to QCs has been found so far.

In the tight-binding model of some QCs, where electrons hop from site to site along the edge of a tile in the quasiperiodic tiling, %
local geometry of the quasicrystalline lattice
can give rise to strictly localized states of electrons. 
Such localized %
or confined states are highly degenerate and result in a large peak at zero energy in the density of states (DOS) \cite{Kohmoto86, Arai88, Rieth95, Koga17, Araujo19, Day20, Koga20, Oktel21, Ha21, Koga21, Keskiner22, Ghadimi23, Matsubara24}.
For example, Penrose \cite{Kohmoto86, Arai88, Rieth95, Koga17}, Ammann-Beenker \cite{Araujo19, Koga20, Oktel21}, and Socolar dodecagonal \cite{Socolar89, Koga21, Keskiner22} QCs exhibit a flat band at zero kinetic energy, which arises from a large family of confined states.
\red{If the chemical potential is at or close to zero energy, a flat band of confined states -- when combined with the underlying inhomogeneity of the quasicrystalline lattice -- } may lead to a superconducting phase that has no counterpart in periodic systems.

In this work, we investigate superconductivity in Ammann-Beenker QCs at and near half filling.  %
We demonstrate that this system presents gapless superconductivity under magnetic field, where the bulk energy spectrum is gapless despite the finite superconducting order parameter at all sites.
Gapless superconductivity was originally studied in conventional $s$-wave superconductors with magnetic impurities \cite{Abrikosov60, Woolf65, Balatsky06}.
In such systems, a band is formed inside the superconducting energy gap, which is broadened as the impurity concentration is increased,
leading up to gapless superconductivity \cite{Yu65, Shiba68, Rusinov69}.
In the presence of magnetic field, disordered superconductors can also exhibit gapless superconductivity \cite{Nanguneri12, Jiang13, Babkin23arXiv}.
In contrast to these previous studies, we find that quasicrystalline gapless superconductivity occurs %
without additional disorder or magnetic impurities, and only at and very close to half filling in Ammann-Beenker QCs. %
\red{This phenomenon is unique to QCs in that it originates in the combination of two intrinsic features of QCs, namely, the lack of translational symmetry and %
the highly degenerate flat band of confined states.} %
Furthermore, we show that gapless superconductivity can coexist with nontrivial topology induced by Rashba spin-orbit coupling.
Gapless superconducting phases can be classified topologically in terms of a real-space topological invariant called the pseudospectrum invariant \cite{loring2015k, cerjan2022local}.
The topological characterization differs from conventional approaches that rely on the existence of an energy gap \cite{Chiu_2016}. 
In the gapless topological superconducting phase, %
we find topologically protected edge modes.

\textit{Model.}
We start with studying superconductivity in Ammann-Beenker QCs at half filling. %
Structure of the QC is shown in Supplemental Material \cite{SupplementalMaterial}.
The mean-field Hamiltonian for an $s$-wave superconductor with Rashba spin-orbit coupling and Zeeman field \cite{Sato09,Sato10} can be written in the Bogoliubov de-Gennes (BdG) formalism as 
\begin{equation}
    H=\frac{1}{2}\sum _{ij\alpha \beta}
    (c^{\dagger}_{i\alpha}~ c_{i\alpha})
    \mathcal{H}_{\mathrm{BdG}}
    \begin{pmatrix}
        c_{j\beta} \\ c_{j\beta}^{\dagger}
    \end{pmatrix},\;
    \mathcal{H}_{\mathrm{BdG}}=
    \begin{pmatrix}
        \mathcal{H} & \bm{\Delta} \\
        \bm{\Delta} ^{\dagger} & -\mathcal{H}^{\ast}
    \end{pmatrix},    
    \label{M-eq: Hamiltonian}
\end{equation}
with $c^{\dagger}_{i\alpha}$ the creation operator of an electron with spin $\alpha$ at site $i$, which corresponds to the $i$th vertex of the Ammann-Beenker tiling.
The single-particle Hamiltonian $\mathcal{H}$ is given by \cite{Goertzen17,Hori24arXiv}
\begin{align}
    \mathcal{H}_{i\alpha, j\beta} 
    &= \bigl[t_{ij}\sigma_0 - \bigl(\tilde{\mu}-V^{(H)}_{i\bar{\alpha}}+\bar{V}^{(H)}_{\bar{\alpha}}\bigr)\delta _{ij}\sigma_0
    - \tilde{h} \delta _{ij} \sigma_3 \notag \\ 
    &+ \imath \lambda _{\mathrm{R}} (\sigma_1 y_{ij} - \sigma_2 x_{ij})\bigr]_{\alpha \beta},
    \quad\bar{\alpha}\ne \alpha,
    \label{M-eq: H matrix}
\end{align}
where $\bm{\sigma} = (\sigma_1, \sigma_2, \sigma_3)$ are the Pauli matrices acting on spin space, $\sigma_0$ the $2 \times 2$ identity matrix, and $\imath$ the imaginary unit $\sqrt{-1}$.
$t_{ij}\equiv -t$ ($t>0$) is the hopping amplitude from the $j$th to $i$th site,
$\lambda_{\mathrm{R}}$ is the Rashba spin-orbit coupling constant, $x_{ij}$ ($y_{ij}$) is the $x$ ($y$) component of the unit vector connecting site $j$ to site $i$, and 
\begin{equation}
    V^{(H)}_{i\alpha} = U \langle c_{i\alpha}^\dagger c_{i\alpha} \rangle,\quad 
    \bar{V}^{(H)}_{\alpha} = \frac{1}{N}\sum_i V^{(H)}_{i\alpha},
    \label{Hartree}
\end{equation}
where $U<0$ is the coupling constant of an onsite pairing interaction and $N$ is the number of lattice sites in the system.
$\tilde{\mu}=\mu-(\bar{V}^{(H)}_{\uparrow}+\bar{V}^{(H)}_{\downarrow})/2$ and 
$\tilde{h}=h_z+(\bar{V}^{(H)}_{\uparrow}-\bar{V}^{(H)}_{\downarrow})/2$, with the chemical potential $\mu$ and the Zeeman energy $h_z$.
The off-diagonal elements in Eq.~(\ref{M-eq: Hamiltonian}) are given by
\begin{align}
    \bm{\Delta}_{i \alpha, j\beta }=[\imath \Delta _i\delta _{ij}\sigma _2]_{\alpha \beta}\,,\quad \Delta_i = U \langle c_{i\downarrow} c_{i\uparrow} \rangle.
    \label{OrderP}
\end{align}
We solve the BdG equations self-consistently for the Hartree potential and the superconducting order parameter in Eqs.~(\ref{Hartree}) and (\ref{OrderP}), respectively.
Below we present results for $U=-1.5t$ and $h_z \leq 0$ at zero temperature.
In the following, we use an approximant (a quasicrystal) with $N=2786$ ($N=2869$) with the periodic (open) boundary condition, unless $N$ is specified otherwise.

\begin{figure}[t]
    \centering
    \includegraphics[scale=0.14]{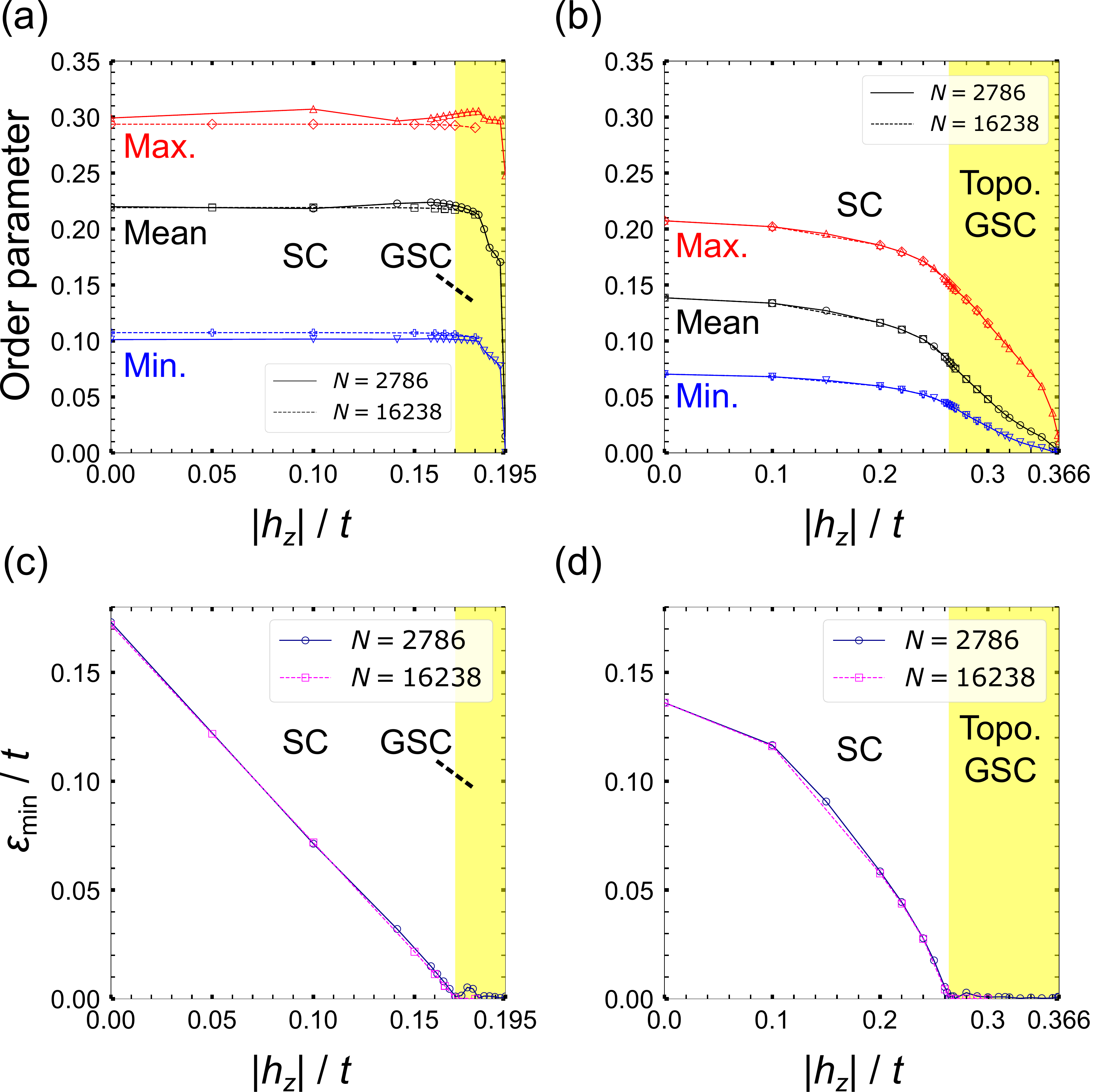}
    \caption{
    (a) and (b) Maxmimum, mean, and minimum absolute values of the superconducting order parameter scaled by $t$ as a function of $|h_z|/t$ for $\lambda_{\mathrm{R}}=0$ and $0.5t$, respectively, in the Ammann-Beenker QC with the periodic boundary condition. SC (GSC) indicates gapful (gapless) superconducting phase.
    (c) and (d) Lowest absolute excitation energy $\varepsilon _{\mathrm{min}}/t$ as a function of $|h_z|/t$ for $\lambda_{\mathrm{R}}=0$ and $0.5t$, respectively. %
    The results for $N=16\,238$ are also shown.
    In (b) and (d), the gapless superconducting phase is topologically nontrivial.
    The finite gap near $|h_z|/t\simeq 0.18t$ in (c) for $N=2786$ is an artifact due to relatively small size.
    }
    \label{fig:MH_fig2}
\end{figure}

\textit{Gapless superconductivity and confined states.}
Using the model for superconducting Ammann-Beenker QCs in Eq.~(\ref{M-eq: Hamiltonian}), we examine the bulk energy spectrum with the periodic boundary condition.
Figure~\ref{fig:MH_fig2}(a) [\ref{fig:MH_fig2}(b)] shows the magnitude of the converged order parameter in the absence (presence) of Rashba spin-orbit coupling as a function of $|h_z|/t$, along with the result for a larger system with $N=16\,238$.
The red, black, and blue data are, respectively, the maximum, mean, and minimum values of $\{|\Delta_i|\}$.
The mean value is defined as $\bar{\Delta } =(1/N)\sum _{i}|\Delta_i|$. %
Superconductivity is destroyed at $h_z \simeq -0.195 t$ ($-0.366t$) when $\lambda _{\mathrm{R}}= 0$ ($0.5t$), with the order parameter vanishing at most (all) sites.

Figures~\ref{fig:MH_fig2}(c) and \ref{fig:MH_fig2}(d) present the lowest absolute energy $\varepsilon _{\mathrm{min}}$ of quasiparticle excitation for %
$\lambda _{\mathrm{R}} =0$ and $0.5t$, respectively.
In both cases, the bulk energy gap is suppressed as $|h_z|/t$ increases. The bulk gap closes at 
$h_z \simeq -0.17t ~(-0.265t)$ for $\lambda _{\mathrm{R}}= 0 ~(0.5t)$, while the superconducting order parameter remains finite at all sites. 
We have thus found a stable gapless superconducting phase in Ammann-Beenker QCs at half filling under magnetic field.
This quasicrystalline gapless superconductivity is qualitatively different from that in disordered systems, in which strong impurities %
make the order parameter vanish at some lattice sites \cite{Ghosal01,Nanguneri12,Jiang13}.
We have also confirmed the occurrence of gapless superconductivity in the system with $N=16\,238$ as %
can be seen in Fig.~\ref{fig:MH_fig2}(c) and \ref{fig:MH_fig2}(d).
Moreover, we show in Supplemental Material \cite{SupplementalMaterial} that gapless superconductivity appears even when the system is not exactly at half filling.

\begin{figure*}[t]
    \includegraphics[width=2.\columnwidth]{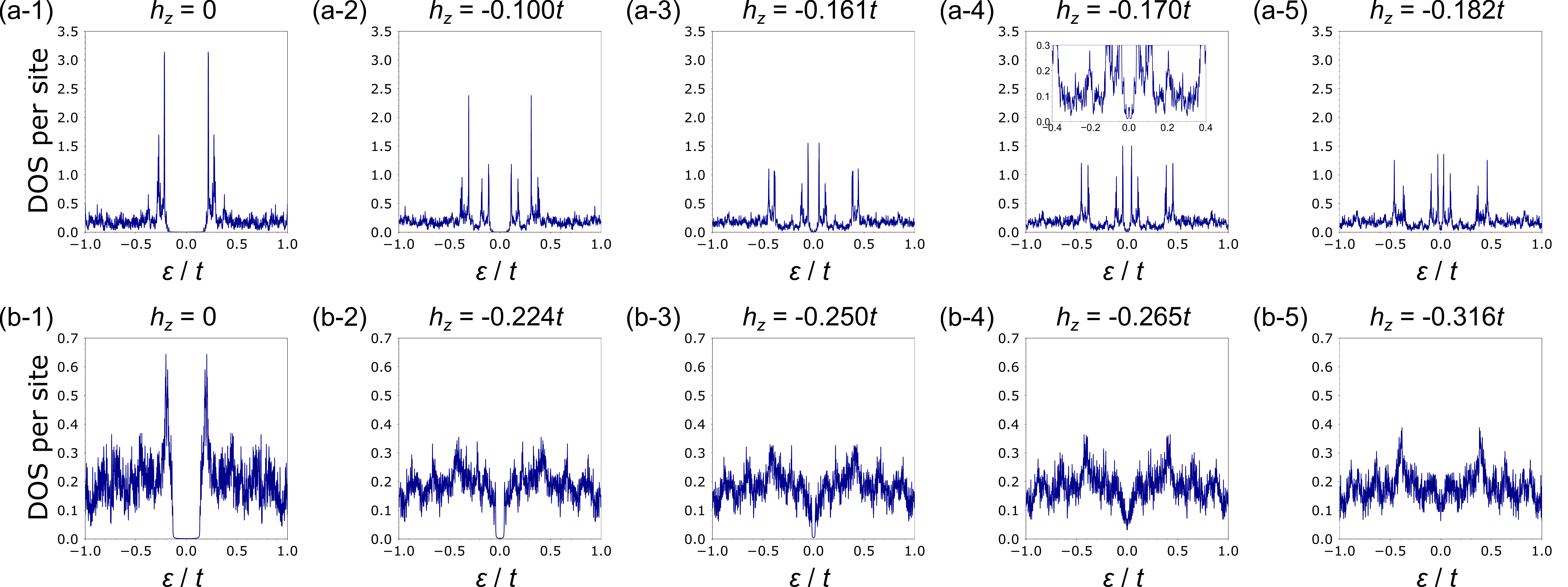}
    \caption{
    Dependence of the DOS of the superconducting Ammann-Beenker QC on 
    $h_z/t$ for $\lambda _{\mathrm{R}}=0$ [(a-1)-(a-5)] and $0.5t$ [(b-1)-(b-5)]
    with the periodic boundary condition. The spectral gap closes at $h_z \simeq -0.170t$ and $-0.265t$ for $\lambda _{\mathrm{R}}= 0$ and $0.5t$, respectively, and the
    DOS in (a-4), (a-5), (b-4), and (b-5) shows gapless superconductivity. A magnified view is given in the inset in (a-4). 
    }
    \label{fig:DOS}
\end{figure*}

We examine the DOS of the superconducting Ammann-Beenker QC to understand the mechanism of gapless superconductivity.
Presented in Fig.~\ref{fig:DOS}(a-1)-\ref{fig:DOS}(a-5) is the DOS for $\lambda_{\mathrm{R}}=0$ and varying $h_z/t$.
In the absence of magnetic field [Fig.~\ref{fig:DOS}(a-1)],
the superconductor has $\bar{\Delta }_0\equiv \bar{\Delta }(h_z=0)\simeq 0.22t$ [Fig.~\ref{fig:MH_fig2}(a)] and $\varepsilon _{0}\equiv \varepsilon _{\mathrm{min}}(h_z=0) \simeq 0.173t$  [Fig.~\ref{fig:MH_fig2}(c)].
Without spin-orbit coupling, we need a Zeeman energy comparable to the bulk %
superconducting gap $\varepsilon _{0}$ for realizing gapless superconductivity \cite{Kinnunen18}.
In periodic crystals, %
such strong Zeeman field would destroy conventional superconductivity as $|h_z| \geq \varepsilon _{0} = \bar{\Delta }_0$, which exceeds the Chandrasekhar-Clogston limit \cite{Chandrasekhar62, Clogston62}.
On the contrary, Ammann-Beenker QCs at half filling exhibit gapless superconductivity under magnetic field. %
With nonzero Zeeman field, the twofold degeneracy of each orbital state is lifted and 
quasiparticle states near the gap edges approach zero energy as $|h_z|$ increases, reducing the energy gap %
[Figs.~\ref{fig:DOS}(a-2) and \ref{fig:DOS}(a-3)].
The gap closes at $|h_z| \simeq 0.17t \simeq \varepsilon _0$ [see the inset in Fig.~\ref{fig:DOS}(a-4)]
and a gapless superconducting phase emerges [Figs.~\ref{fig:DOS}(a-4) and \ref{fig:DOS}(a-5)]. 

Gapless superconductivity can be realized in Ammann-Beenker QCs because a large number of confined states and broken translational symmetry allow the QC to satisfy $\varepsilon _{0} < \bar{\Delta }_0$, a favorable condition for gapless superconductivity to occur.
First, at or close to half filling, %
$\bar{\Delta }_0$ is strongly enhanced due to highly degenerate confined states %
in the normal state
and is much larger than that at low or high filling.
The strong degeneracy of confined states results in the sharp coherence peaks as can be seen in Fig.~\ref{fig:DOS}(a-1).
Second, as also visible in Fig.~\ref{fig:DOS}(a-1), the gap-edge singularity is smeared by the underlying inhomogeneity of the quasicrystalline lattice, i.e., lowest-energy quasiparticle states spread towards zero energy in the DOS \cite{Balatsky06, Ghosal98, Ghosal01}.
These two factors combine to yield $\varepsilon _{0} < \bar{\Delta }_0$ in Ammann-Beenker QCs at half filling. This mechanism of gapless superconductivity originating in confined states and the lack of translational symmetry is further elucidated in Supplemental Material \cite{SupplementalMaterial}, where it is also shown that gapless superconductivity disappears as one moves away from half filling when $\varepsilon _0 \simeq \bar{\Delta} _0$.

When Rashba spin-orbit coupling is present, confined states are all destroyed. %
It can be seen in Fig.~\ref{fig:DOS}(b-1)-\ref{fig:DOS}(b-5) for $\lambda_{\mathrm{R}}=0.5t$ (note the reduced scale for the DOS) that the coherence peaks are reduced drastically for $h_z=0$ and %
almost nonexistent for $h_z\ne 0$. %
Even though $\varepsilon_0 \simeq 0.136t$ is smaller than that for $\lambda_{\mathrm{R}}=0$ and the bulk gap decreases quickly as $|h_z|$ increases [Figs.~\ref{fig:DOS}(b-2) and \ref{fig:DOS}(b-3)], stronger Zeeman field is required for gapless superconductivity to occur at $|h_z|\simeq 0.265t$ [Figs.~\ref{fig:DOS}(b-4) and \ref{fig:DOS}(b-5)]. %

\textit{Topology of gapless superconductivity.}
Spin-orbit coupling can lead to various topological phases in crystals %
that can host topological boundary modes \cite{Hasan10, Qi11}.
Motivated by this, we investigate whether Rashba spin-orbit coupling can induce nontrivial topology in gapless superconductivity in Ammann-Beenker QCs --
despite the lack of a bulk energy gap as well as translational symmetry.
To explore topological properties,
we utilize a spectral localizer \cite{loring2015k, cerjan2022local, cerjan2023quadratic}.
Topological classification based on the spectral localizer is applicable to gapless and aperiodic systems.
The model in Eq.~(\ref{M-eq: Hamiltonian}) belongs to class D in the Altland-Zirnbauer classification
since it describes a superconductor %
with broken time-reversal symmetry due to 
the Zeeman field \cite{Altland_1997}.
For class D in two dimensions, the spectral localizer $L(X, Y, \mathcal{H}_\mathrm{BdG})$ is defined as \cite{loring2015k, Fulga16, cerjan2022local}
\begin{align}
    \begin{split}
    &L_{(x,y)}(X,Y,\mathcal{H}_{\mathrm{BdG}}) \\
    &=\begin{pmatrix}
        \kappa (X-xI) & \kappa (Y-yI)-\imath \mathcal{H}_{\mathrm{BdG}} \\
        \kappa (Y-yI)+\imath \mathcal{H}_{\mathrm{BdG}} & -\kappa (X-xI)
    \end{pmatrix},    
    \end{split}
    \label{eq:spectralLocalizer}
\end{align}
where $(x,y)$ is the coordinate, %
$X$ ($Y$) is the position operator for the $x$ ($y$) component of a lattice site, %
and $I$ is the identity matrix.
Here, $\kappa > 0$ is a scaling coefficient to compensate the different dimensions %
of the position operators and the Hamiltonian. %
In terms of the spectral localizer in Eq.~(\ref{eq:spectralLocalizer}), 
a topological invariant called a pseudospectrum invariant $C_{\mathrm{L}}(x, y)$ is defined by \cite{loring2015k, cerjan2022local, cerjan2023quadratic} 
\begin{align}
    \begin{split}
        &C_{\mathrm{L}}(x, y) = \frac{1}{2} \mathrm{sig}[L_{(x, y)}(X, Y, \mathcal{H}_{\mathrm{BdG}})],
    \end{split}
    \label{M-eq: local Chern number}
\end{align}
where $\mathrm{sig}[L_{(x,y)}]$ denotes the difference between the numbers of positive and negative eigenvalues of the spectral localizer $L_{(x,y)}$.
The system is topologically nontrivial if the pseudospectrum invariant is nonzero in some areas of the system in real space. 

\begin{figure}[t]
    \centering
    \includegraphics[scale=0.13]{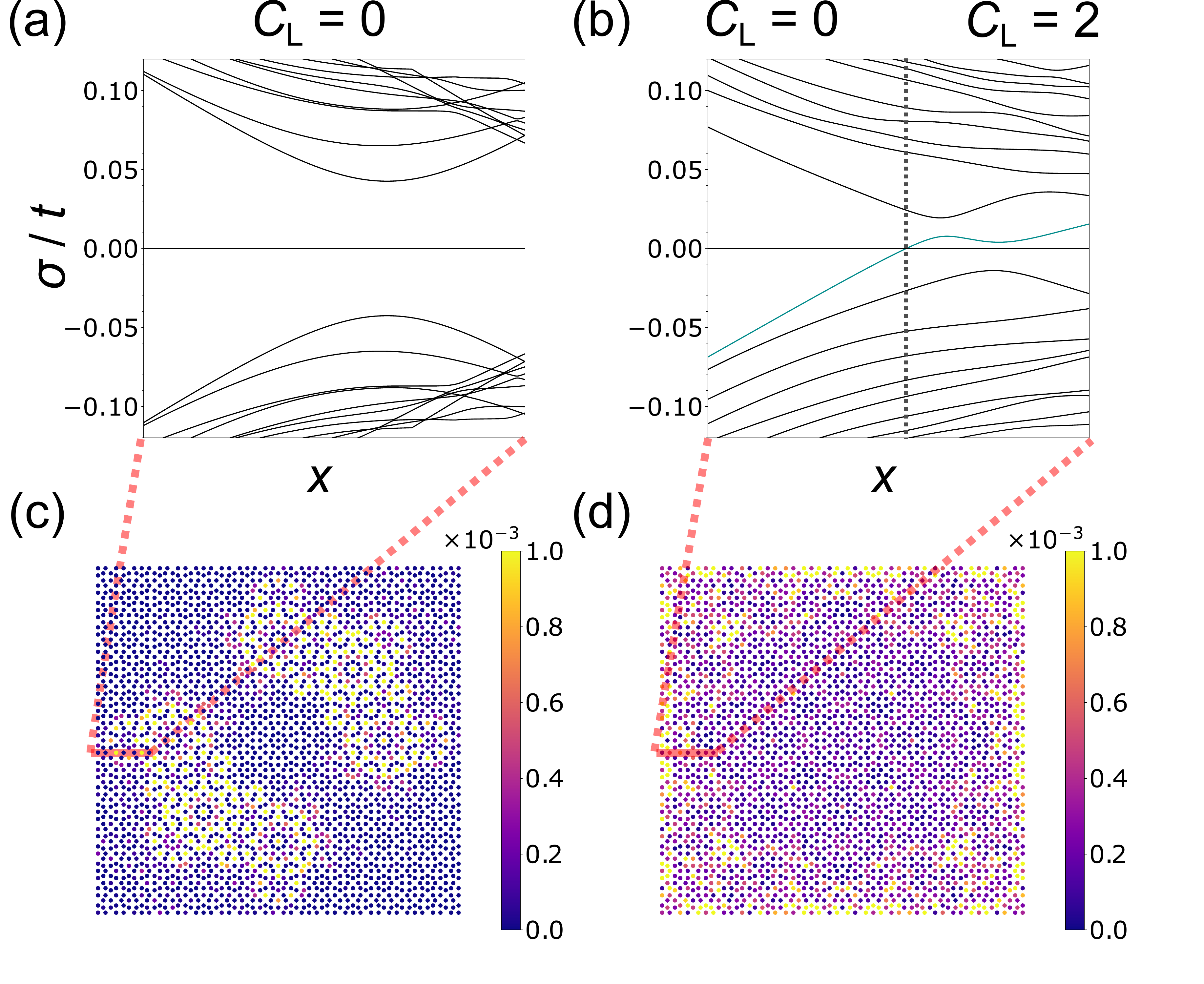}
    \caption{
    Eigenvalue Spectra $\sigma (L_{x,y})$ of the spectral localizer as a function of $x$ along the red-dotted path ($y=0$) in (c) and (d), respectively, for (a) $\lambda _{\mathrm{R}}=0$ and $h_z=-0.182t$ and (b) $\lambda _{\mathrm{R}}=0.5t$ and $h_z=-0.316t$.
    In (b), the eigenvalue that crosses zero is highlighted in cyan.
    (c) and (d) Probability distribution of a bulk mode and an edge mode, respectively, in the Ammann-Beenker QC with the open boundary condition.  
    }
    \label{fig:MH_fig3}
\end{figure}

We have calculated the eigenvalue spectrum $\sigma (L_{(x,y)})$ of the spectral localizer in the Ammann-Beenker QC presenting gapless superconductivity, with the open boundary condition.
Figure~\ref{fig:MH_fig3}(a) shows the spectrum $\sigma (L_{(x,y)})$ in units of $t$ as a function of $x$ along the red-dotted path ($y=0$) in Fig.~\ref{fig:MH_fig3}(c) for $\lambda _{\mathrm{R}}=0$ and $h_z=-0.182t$.
$\kappa=0.02t$ was used in Eq.~(\ref{eq:spectralLocalizer}) for this calculation.
We find that without Rashba spin-orbit coupling,
the superconducting system has no area where the pseudospectrum invariant is nonzero, irrespective of the magnetic field.
Namely, the system %
is topologically trivial before and after the bulk gap closing.

In contrast, when spin-orbit coupling is present,
gapless superconductivity can be topologically nontrivial.
The nontrivial topology can be confirmed by %
$C_{\mathrm{L}}(x, y)$ in Eq.~(\ref{M-eq: local Chern number}) being nonzero.
In Fig.~\ref{fig:MH_fig3}(b) $\sigma (L_{(x,y)})/t$ is plotted as a function of $x$ along the red-dotted path in Fig.~\ref{fig:MH_fig3}(d) (the same path as for Fig.~\ref{fig:MH_fig3}(a)) for $\lambda _{\mathrm{R}}= 0.5t$ and $h_z=-0.316t$.
As $x$ varies from the edge towards the bulk,  
$C_{\mathrm{L}}(x,0)$ changes from 0 to 2 as a doubly degenerate eigenvalue of the spectral localizer %
crosses zero [Fig.~\ref{fig:MH_fig3}(b)].
Therefore, %
this gapless phase is topologically nontrivial.
In Supplemental Material \cite{SupplementalMaterial},
we discuss topological phase transitions for gapless superconductivity in detail.
We note that the same model in Eq.~(\ref{M-eq: Hamiltonian}) on a square lattice %
does not exhibit gapless 
topological superconductivity in the presence of Rashba spin-orbit coupling \cite{Sato09, Sato10}.

The gap of the spectral localizer closing indicates the existence of edge modes associated with a nonzero pseudospectrum invariant 
in the bulk, regardless of whether they are hybridized with bulk states \cite{cerjan2022local}. 
In a gapless system with continuous spectrum across zero energy, edge modes are expected to hybridize with bulk states, with their energy lifted from zero as a result.
In the absence of Rashba spin-obrit coupling, 
we have not found any edge state 
in the gapless superconducting phase. The probability distribution of a bulk state with energy $\varepsilon \simeq 9\times 10^{-5}t$ is shown for $h_z=-0.182t$ in Fig.~\ref{fig:MH_fig3}(c).
With nontrivial topology induced by Rashba spin-obrit coupling, edge states do appear with near-zero energy.
This is illustrated for $\lambda _{\mathrm{R}}= 0.5t$ and $h_z=-0.316t$ in Fig.~\ref{fig:MH_fig3}(d), where the probability distribution of an edge state with energy $\varepsilon \simeq 3.2 \times 10^{-3}t$ is shown.

The pseudospectrum invariant is equivalent to the Chern number in systems with a bulk energy gap \cite{loring2020spectral, Lozano19}.
In a topological gapless superconductor, if the bulk spectrum can be made to have a gap about zero energy while $C_{\mathrm{L}}(x, y)$ remains nonzero in the bulk, the edge modes turn out to be Majorana zero modes. Therefore, the edge state presented in Fig.~\ref{fig:MH_fig3}(d) can be regarded as a Majorana mode buried in the bulk spectrum.

\textit{Conclusion.}
In this Letter, we have demonstrated gapless superconductivity in Ammann-Beenker quasicrystals at half filling under magnetic field.
We have revealed that gapless superconductivity stems from interplay of broken translational symmetry and highly degenerate confined states, \red{which are both} characteristic of quasicrystals.
In the presence of Rashba spin-orbit coupling, the gapless superconducting phase can be topologically nontrivial.
The nontrivial topology can be identified by a nonzero pseudospectrum invariant given by a spectral localizer 
applicable to 
gapless and aperiodic systems.
The quasicrystalline gapless superconducting phase hosts edge modes protected by a nonzero pseudospectrum invariant in the bulk.
Our findings on Ammann-Beenker quasicrystals are expected to hold in other quasicrystals that exhibit a flat band of confined states \cite{Kohmoto86, Arai88, Rieth95, Koga17, Socolar89, Koga21, Keskiner22}.
Quasicrystals can be a unique platform for realizing gapless superconductivity with nontrivial topology in the bulk, presenting topological edge modes.

This work was supported by JST SPRING (Grant No. JPMJSP2151), JSPS KAKENHI (Grants No. JP23K13033, No. JP24K00586, and No. 25H01248), and the Natural Sciences and Engineering Research Council of Canada. This research was made feasible in part by support provided by the Digital Research Alliance of Canada (\href{https://alliancecan.ca/en}{alliancecan.ca}).

\bibliography{QGTSC}

\newpage

\begin{center}
    \bf\large{Supplemental Material}
\end{center}

\begin{figure}[b]
    \centering
    \includegraphics[scale=0.32]{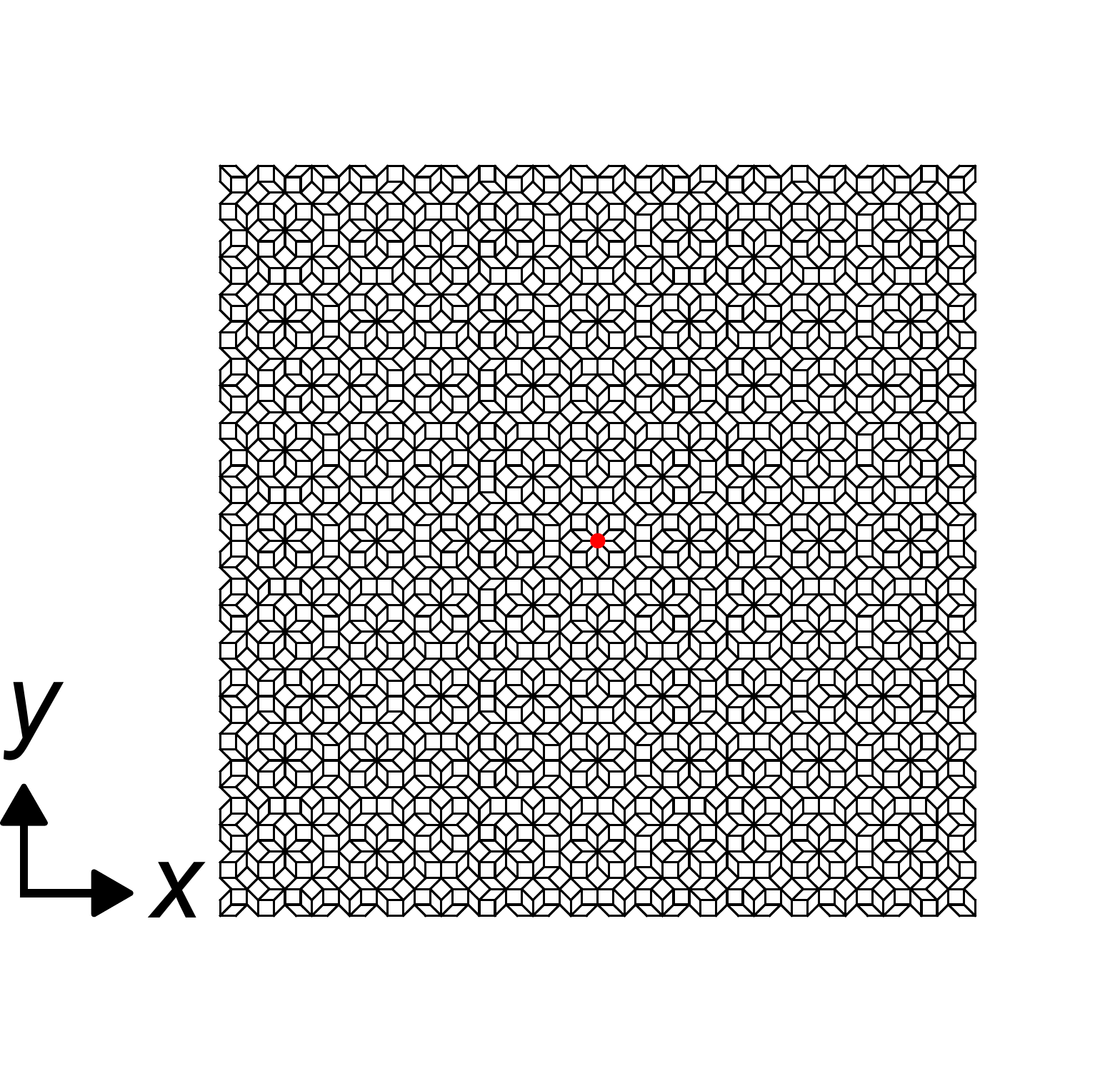}
    \caption{
    Ammann-Beenker QC with 2869 sites. The red point indicates the center of this QC.
    }
    \label{fig:MH_fig1}
\end{figure}

\begin{figure}[b]
    \centering
    \includegraphics[width=\columnwidth]{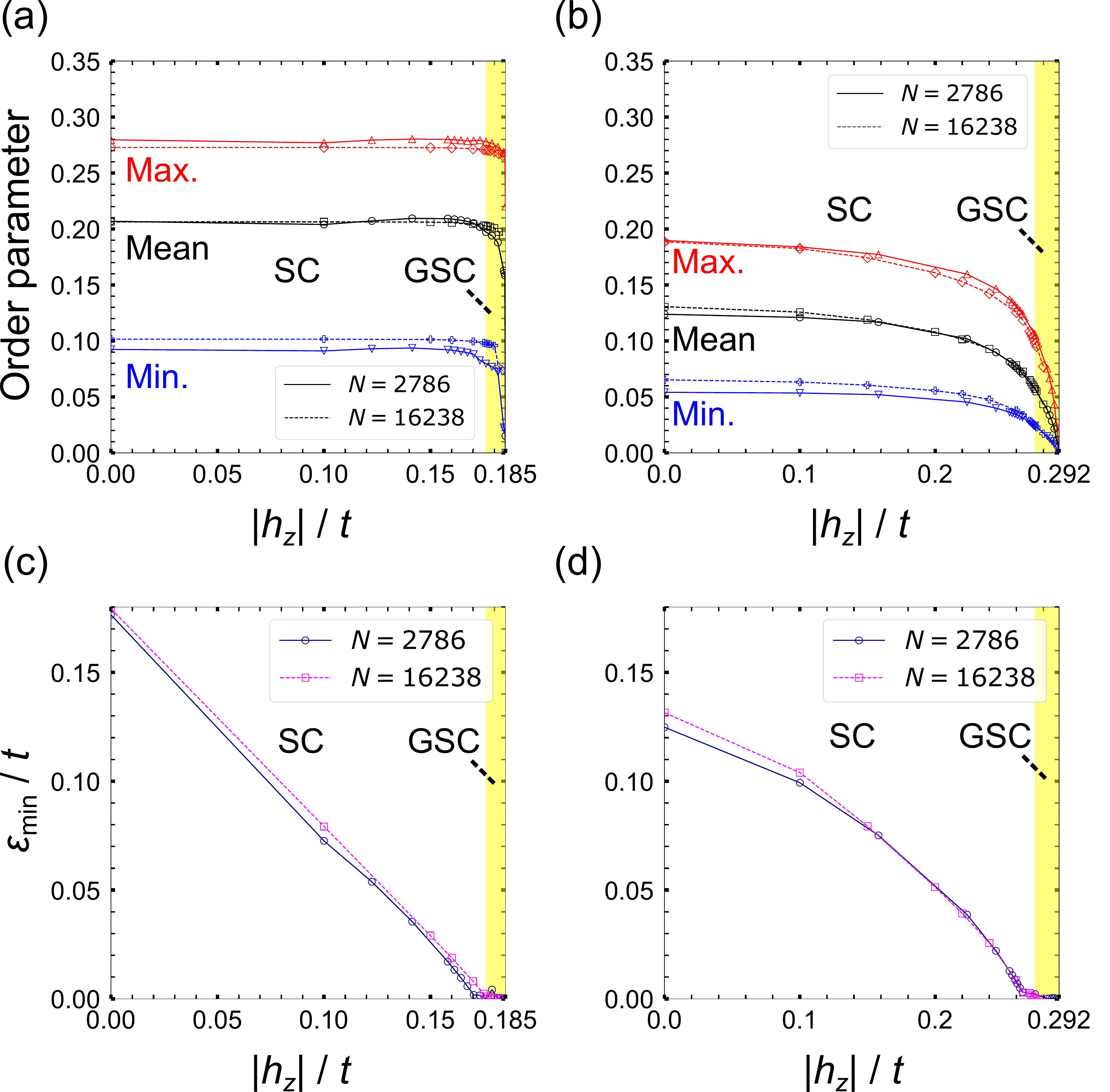}
    \caption{(a,b) Maximum, mean, and minimum absolute value of the superconducting order parameter scaled by $t$ and (c,d) lowest absolute excitation energy $\varepsilon_{\mathrm{min}} / t$
    as a function of $|h_z|/t$ for (a,c) $\lambda_{\mathrm{R}} = 0$ and (b,d) $\lambda_{\mathrm{R}}=0.5t$ in the 2786- and 16238-site Ammann-Beenker QCs with the periodic boundary condition. 
    SC and GSC indicate gapful and gapless superconducting phases, respectively.
    }
    \label{fig: gsc near half filling}
\end{figure}

\section{I. Structure of Ammann-Beenker quasicrystals}
We illustrate the structure of Ammann-Beenker quasicrystals (QCs)
in Fig.~\ref{fig:MH_fig1}. 
This figure presents the Ammann-Beenker QC with 2869 sites, 
which we have used for calculations with the open boundary condition.
The same QC is shown 
in Figs.~3(c) and 3(d) of the Letter. 
We have used the method of Ref.~\cite{Ghadimi21} based on the inflation/deflation rules to generate Ammann-Beenker QCs and their approximants, e.g., with 2869 and 2786 sites, respectively. The method of producing approximants determines the allowed values of the number of lattice sites $N$ in an approximant.
In the conventional method based on projection of a hypercubic lattice in higher-dimensional space, $N$ jumps from 1393 to 8119, where $N=1393$ is too small to see bulk effects, while $N=8119$ is numerically too costly to perform a large number of self-consistent calculations, e.g., to make a phase diagram.
We find that $N=2786$ is large enough to study bulk physics, yet so small that a series of self-consistent calculations can readily be completed. With the same seed as used in Ref.~\cite{Ghadimi21}, there is a substantial jump between $N=2786$ and the next generation with $N=16\,238$. However, using a different seed in this method can generate an approximant with $N=5572$, for example.

\begin{figure*}[t]
    \includegraphics[width=2.\columnwidth]{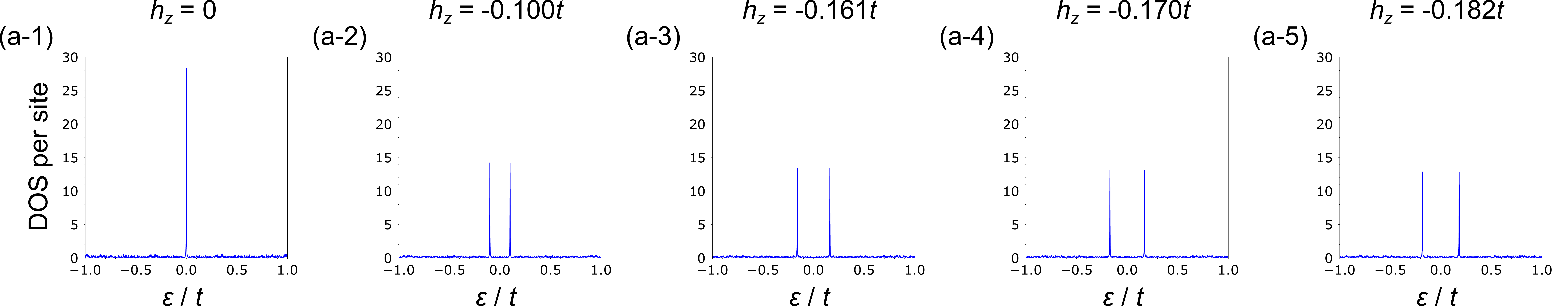}
    \caption{
    Dependence of the DOS in the normal state of the 2786-site Ammann-Beenker QC with the periodic boundary condition on the Zeeman energy $h_z$ for $\lambda _{\mathrm{R}}=0$.
    }
    \label{fig:NS-DOS}
\end{figure*}

\section{II. Gapless superconductivity near half filling}
We have confirmed that gapless superconductivity emerges in Ammann-Beenker QCs slightly away from half filling as well.
Figure~\ref{fig: gsc near half filling}(a) [\ref{fig: gsc near half filling}(b)] shows the magnitude of the superconducting order parameter in the absence (presence) of Rashba spin-orbit coupling for $\tilde{\mu} = -0.1t$, $U = -1.5t$, and $\lambda_{\mathrm{R}}=0$ ($0.5t$) as a function of the Zeeman energy $|h_z|/t$ in the 2786- and 16238-site 
Ammann-Beenker QCs with the periodic boundary condition.
The red, black, and blue data represent the maximum, mean, and minimum values, respectively, of the 
self-consistently obtained 
$\{ |\Delta _i| \}$.
Superconductivity is destroyed at $h_z \simeq -0.185t$ ($-0.292t$) when $\lambda_{\mathrm{R}} = 0$ ($0.5t$).

In Fig.~\ref{fig: gsc near half filling}(c) and \ref{fig: gsc near half filling}(d), the lowest absolute excitation energy is plotted as a function of $|h_z|/t$ in the respective systems with $\lambda_{\mathrm{R}}=0$ and $\lambda_{\mathrm{R}}=0.5t$.
In both cases, the spectral gap decreases as the magnetic field increases.
For $\lambda_{\mathrm{R}} = 0~(0.5t)$ the bulk gap closes at $h_z \simeq -0.177t~(-0.274t)$ in the 16238-site Ammann-Beenker QC,
while the converged superconducting order parameter is finite at all sites.
Thus, a stable gapless superconducting phase appears.
Unlike at half filling, however, it is limited to a very narrow region of $|h_z|$ of roughly $0.01t$ ($0.02t$) for $\lambda_{\mathrm{R}} = 0~(0.5t)$. In the presence of Rashba spin-orbit coupling, 
we have not found topological gapless superconductivity for nonzero $\tilde{\mu}$.

\section{III. Details of the mechanism of quasicrystalline gapless superconductivity}
The density of states (DOS) in the normal state of the 2786-site Ammann-Beenker QC with the periodic boundary condition in the absence of Rashba spin-orbit coupling 
is presented in Fig.~\ref{fig:NS-DOS}(a-1)-\ref{fig:NS-DOS}(a-5) 
for various values of $h_z$. 
It can be seen in Fig.~\ref{fig:NS-DOS}(a-1) 
that the large number of confined states results in a very high peak at zero energy in the normal-state DOS.
When the magnetic field is applied, each orbital state is split with the energy shift $\pm h_z$ and so is the zero-energy peak in Fig.~\ref{fig:NS-DOS}(a-1).
[Fig.~\ref{fig:NS-DOS}(a-2)-\ref{fig:NS-DOS}(a-5)].
These two peaks in the normal-state DOS lead to the sharp coherence peaks near the gap edges in the superconducting state when the chemical potential is at or close to zero energy, 
for relatively small Zeeman splitting. 

\begin{figure*}[t]
    \includegraphics[width=2.\columnwidth]{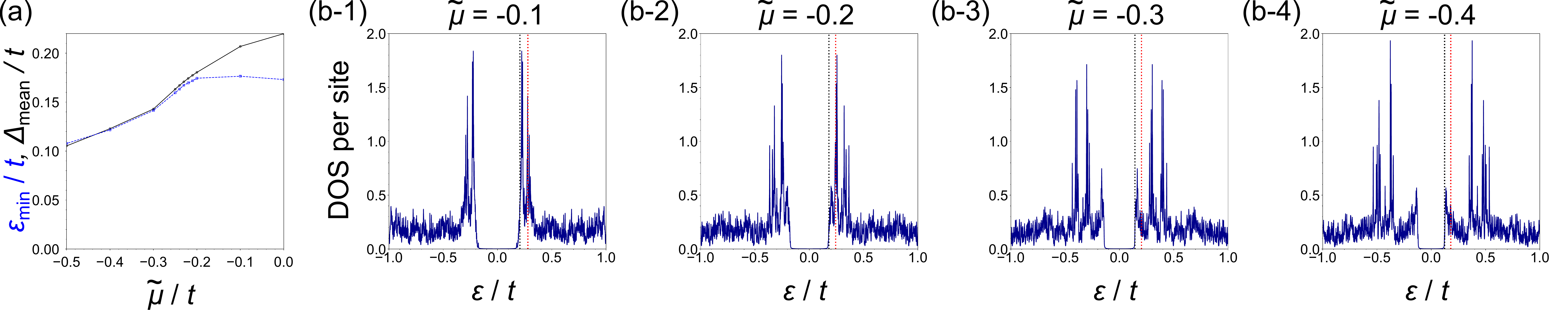}
    \caption{
    (a) Dependence of the bulk spectral gap and the mean value of the superconducting order parameter on the chemical potential $\tilde{\mu}$ in the absence of magnetic field and Rashba spin-orbit coupling. (b-1)-(b-4) DOS in the superconducting Ammann-Beenker QC for various values of $\tilde{\mu}$. The black (red) dotted lines in (b-1)-(b-4) indicate the mean (maximum) value of the order parameter.
    }
    \label{fig:tail states}
\end{figure*}

To understand the mechanism of quasicrystalline gapless superconductivity,
we consider the effects of confined states and inhomogeneity on the lowest absolute excitation energy $\varepsilon _0$ and the mean value $\bar{\Delta} _0$ of the superconducting order parameter $\{|\Delta_i|\}$ in Ammann-Beenker QCs without Zeeman field or spin-orbit coupling.
Since confined states lie at zero energy, 
we have evaluated $\varepsilon _0$ and $\bar{\Delta} _0$ by moving away from half filling, i.e., shifting the chemical potential $\tilde{\mu}$ from zero.
We have found that gapless superconductivity emerges in the presence of magnetic field when $|\tilde {\mu}| \leq 0.2t$,
where $\varepsilon _{0} < \bar{\Delta }_0$, as shown in Fig.~\ref{fig:tail states}(a).

To clarify the roles of confined states and the underlying inhomogeneity of the quasicrystalline lattice on superconductivity, we illustrate how the DOS changes as $\tilde{\mu}$ moves away from zero in Figs.~\ref{fig:tail states}(b-1)-\ref{fig:tail states}(b-4). In these figures, the black and red dotted lines indicate the mean and maximum values of the superconducting order parameter, $\bar{\Delta}_0$ and $\max \{ |\Delta _i| \}$, respectively.
When $|\tilde {\mu}| \leq 0.1t$,
the coherence peaks are sharp due to highly degenerate confined states near the Fermi energy in the normal state [Fig.~\ref{fig:tail states}(b-1)] 
(see also Fig.~2(a-1) in the Letter for $\tilde{\mu}=0$).
This large number of states close to the Fermi energy also enhances the superconducting order parameter. 
At the same time, as can also be seen in Fig.~\ref{fig:tail states}(b-1), the coherence peaks are broadened by 
the site-dependent superconducting order parameter in the QC: 
$\bar{\Delta }_0$ 
is between the spectral gap $\varepsilon _0$ and the coherence peak 
and so $\varepsilon _{0} < \bar{\Delta }_0$.
The mechanism of this broadening is similar to that of a tail in the DOS of disordered superconductors with randomly distributed impurities \cite{Ghosal01,Balatsky06}.
The secondary peak in the DOS roughly at $\max \{ |\Delta _i| \}$ also arises from confined states in the normal state.
For $\tilde{\mu}=-0.2t$, quasiparticle states below $\bar{\Delta }_0$ is hardly visible in Fig.~\ref{fig:tail states}(b-2) in the given scale and this is consistent with Fig.~\ref{fig:tail states}(a), where $\varepsilon_0$ can be seen to be approaching $\bar{\Delta }_0$ from below. 
The secondary peak, which is now the highest in the DOS, is again approximately at $\max \{ |\Delta _i| \}$.
We have confirmed the appearance of gapless superconducting phase under magnetic field 
as long as $|\tilde{\mu }| \leq 0.2t$, although the range of $|h_z|$ for this phase becomes narrower as $|\tilde{\mu}|$ approaches $0.2t$.

As $\tilde{\mu}$ moves further away from zero, $\bar{\Delta}_0$ becomes substantially smaller and roughly equal to $\varepsilon_0$ when $|\tilde{\mu}| \geq 0.3t$, as can be seen in Fig.~\ref{fig:tail states}(a). Thus, for $\tilde{\mu}=-0.3t$ and $-0.4t$ [Figs.~\ref{fig:tail states}(b-3) and \ref{fig:tail states}(b-4)], the gap edge coincides with $\bar{\Delta}_0$. Also, the coherence peaks are much smaller compared to $\tilde{\mu}=-0.1t$, as there are much less number of states at and around the Fermi energy.
The highest peaks in the DOS stemming from confined states are now well beyond the energy scale of $\max \{ |\Delta _i| \}$. These states are so far away from the gap edges that they no longer yield gapless superconductivity under magnetic field. 
We have not found any stable gapless superconducting phase under magnetic field when $\varepsilon _{0} \simeq \bar{\Delta }_0$.

\section{IV. Topological phase transition}
\begin{figure}[h]
    \centering
    \includegraphics[width=\columnwidth]{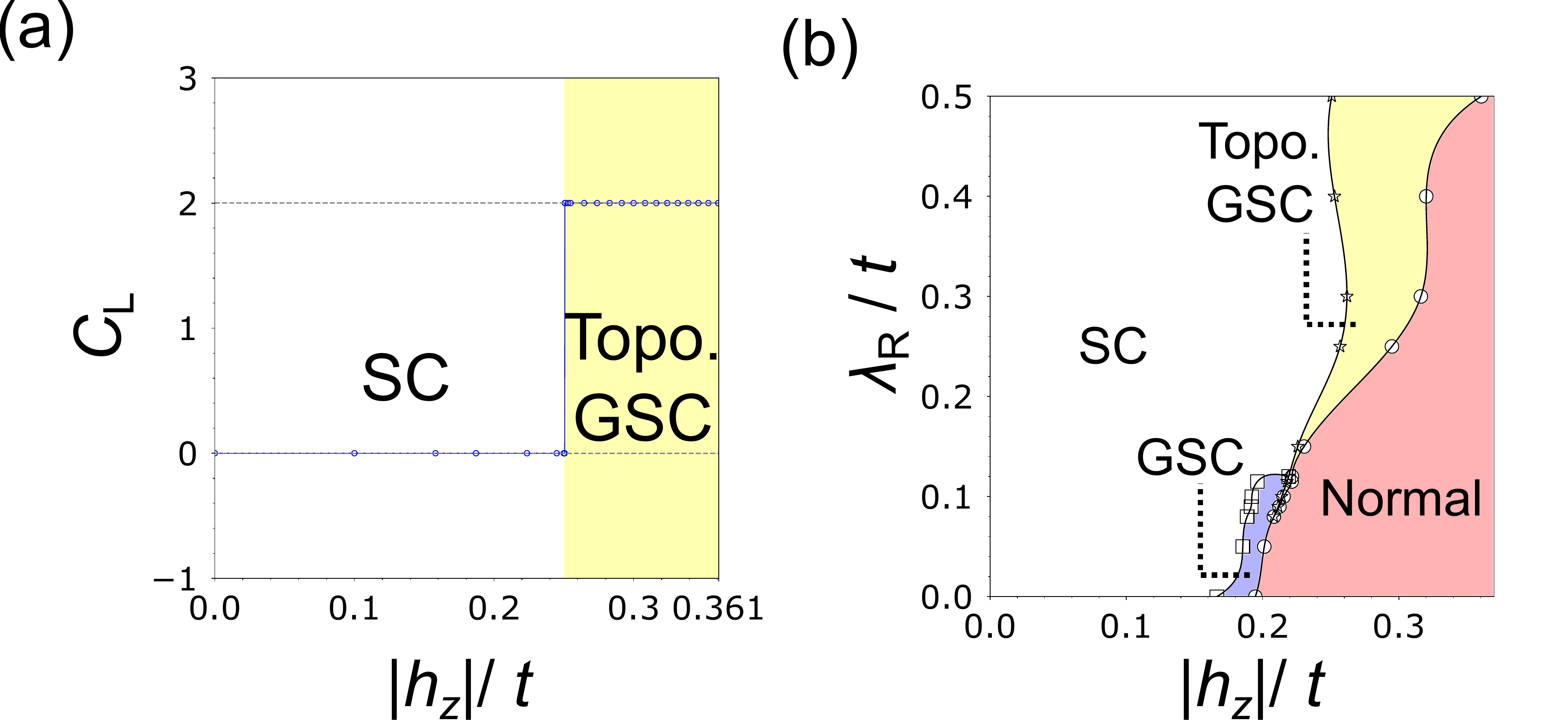}
    \caption{(a) Pseudospectrum invariant at the center of the 2869-site Ammann-Beenker QC with the open boundary condition as a function of the Zeeman energy $|h_z|$ in units of $t$ for $\lambda _{\mathrm{R}}=0.5t$. (b) Topological phase diagram of the same Ammann-Beenker QC in terms of $\lambda _{\mathrm{R}}/t$ and $|h_z|/t$. 
    SC and GSC indicate gapful and gapless superconducting phases, respectively. The topological GSC phase in the yellow region is characterized by $C_{\mathrm{L}} = 2$ in the bulk.
    }
    \label{fig:phasediagram}
\end{figure}

We demonstrate 
topological phase transitions for gapless superconductivity in Ammann-Beenker QCs at half filling.
When Rashba spin-orbit coupling is relatively strong,
nontrivial topology arises simultaneously with the gap closing due to 
magnetic field. 
In Fig.~\ref{fig:phasediagram}(a),
the pseudospectrum index $C_{\mathrm{L}}$ at the center of the Ammann-Beenker QC is plotted as a function of $|h_z|/t$ for $\lambda _{\mathrm{R}}= 0.5t$.
The pseudospectrum invariant changes its value from 0 to 2 at $h_z\simeq -0.251t$.

Figure~\ref{fig:phasediagram}(b) shows a topological phase diagram in terms of $|h_z|/t$ and $\lambda _{\mathrm{R}}/t$.
While gapless superconducting phases appear with weaker Rashba spin-orbit coupling,
the pseudospectrum invariant can be nonzero only when $\lambda_{\mathrm{R}} \geq 0.1t$.

\end{document}